\documentclass[12pt]{JHEP}
\usepackage{epsf,cite}
\newcommand{\be}{\begin{equation}}
\newcommand{\ee}{\end{equation}}
\newcommand{\beqn}{\begin{eqnarray}}
\newcommand{\eeqn}{\end{eqnarray}}

\title{Stable Solitons in Field Theory Models for Tachyon 
Condensation}
\author{Dileep P. Jatkar and Radhika Vathsan\\
 Harish-Chandra Research Institute\thanks{Formerly, Mehta Research
Institute of Mathematics and Mathematical Physics.},\\
 Chhatnag Road, Jhusi, Allahabad 211 019, INDIA\\
Email: \email{dileep@mri.ernet.in} and 
\email{radhika@mri.ernet.in}
}

\abstract{
We study soliton solutions in scalar field theory with a variety of unbounded
potentials. A subset of these potentials have Gaussian lump solutions and
their fluctuation spectrum is governed by the harmonic oscillator problem.
These lumps are unstable with one tachyonic mode. Soliton solutions in several
other classes of potentials are stable and are of kink type.  The problem of
the stability of these solutions is related to a supersymmetric quantum
mechanics problem. The fluctuation spectrum is not equispaced and does not
contain any tachyonic mode. The lowest energy mode is the massless Goldstone
mode which restores broken translation invariance.} 

\keywords{Bosonic Strings, Solitons Monopoles and Instantons}

\preprint{
\hepth{0104229} \\
MRI-P-010401
}

\begin{document}

\section{Introduction}

Tachyon condensation in string theory has attracted considerable attention
following the conjectures of Sen\cite{sen1,sen2,sen3,sen4}. These conjectures
have been studied using string field theory both in the background independent
formulation\cite{bsft1,bsft2,bsft3,bsft4,bsft5} as well as in the cubic string
field theory\cite{csft} using the level truncation scheme\cite{Sam,SZ,MT,KR}.
Strong evidence in support of these conjectures is now
available\cite{HK,KJMT,MSZ,GS,KMM,GhS} due to these results. 

Toy models of tachyon condensation are extremely useful in  learning more
about the dynamics and stability of non-BPS D-branes. In particular, an
exactly solvable toy model, which preserves the basic features of string field
theory, may educate us about the disappearance of the unstable D-brane and
structure of the closed string vacuum. To this end, Zwiebach\cite{Zwie} and
then Minahan and Zwiebach\cite{MZ} have studied some solvable toy models for
tachyon condensation in field theory. While the string field contains an
infinite number fields, which include tachyon, gauge field and higher massive
stringy excitations, their toy model contains only a single scalar field,
representing the tachyon. In the cubic string field theory, the product of
string fields is defined in terms of the $*$-product which involves an
infinite number of derivatives of the field. In their field theory model, the
Lagrangian density contains at most two derivatives of the scalar field.

The model which was studied by Minahan and Zwiebach\cite{MZ} contains a scalar
field with a canonical kinetic term and a potential
\be
V_{\infty}(\phi) = - {1\over 4} \phi^2 \ln(\phi^2),\label{Inf}
\ee
which is obtained by taking $p\to\infty$ limit of
\be
V_{p}(\phi)  =  {p\over 4} \phi^2 (1 - \phi^{2/p} ).\label{finite}
\ee
The model studied by Zwiebach\cite{Zwie}, a particular case of the
potential (\ref{finite}), supports a lump solution, the fluctuation
spectrum of which reveals an unstable (tachyonic) mode. For finite values
of $p$, there are a finite number of discrete fluctuation modes and their
spectrum is not equispaced. However, in the $p\to\infty$ limit, the lump
solution is a Gaussian with the fluctuation spectrum governed by the
Schr\"{o}dinger equation for harmonic oscillator. Consequently, the
fluctuation spectrum is equispaced, which is a feature of the spectrum of
string theory. The potential (\ref{Inf}) is identical to the tachyon
potential in the boundary string field theory. This formulation is
formally background independent. In this case the tachyon mode has mass
$m^2 = -1$ which is another feature common with the string theory tachyon.

In this paper, we enquire if this behaviour of the lump solution and its
fluctuation spectrum are unique to the potential (\ref{Inf}) above or
whether there is some universality in it. In other words, are there other
potentials with some parameter $p$, which in the limit $p\to\infty$, have
a Gaussian lump solution? Do we recover a lump solution whose fluctuation
spectrum is equispaced? We will see that this is not so. We find that a
large class of models do not have unstable lump solutions. Instead, they
have stable kink solutions, despite having unbounded potentials. The
lowest energy mode of fluctuations about these kink solutions is massless.
This is the Goldstone mode of the kink corresponding to broken translation
invariance.

In section II, we will briefly review the work of Minahan and
Zwiebach\cite{MZ}. In section III, we study various classes of potentials,
which include those with a Gaussian lump solution as well as those with more
complicated soliton profiles. Since we want to see if in the limiting case we
recover the Schr\"{o}dinger equation for harmonic oscillator, we will
concentrate on the potentials obtained by taking the $p\to\infty$ limit.  

\section{Review of the Minahan-Zwiebach Potential}

In this section we will briefly review the model of Minahan and
Zwiebach\cite{MZ}. They consider a toy model in field theory to study tachyon
condensation. Unlike open string field theory which contains infinite number
of fields, this model contains only one scalar field. The Lagrangian contains
terms up to two derivatives. In other words, it contains a canonical kinetic
term and a potential term which is a function purely of the scalar field and
not of its derivatives.

The  Lagrangian density is
\be
{\mathcal{L}} = - \frac{1}{2} \partial_{\mu}\phi\partial^{\mu}\phi 
              - V_{p}(\phi),
\ee
where the potential $V_p(\phi)$ is 
\be
V_{p}(\phi)  =  A p \phi^2 (1 - \phi^{2/p} )\label{m-zmodel}.
\ee
The parameter $p$ is an integer and $A$ is chosen to be $1/4$ but we will keep
it as a free parameter. This potential is well defined for $p=1,2$ and for
$p > 2$, $\phi^{2/p} \equiv (\phi^2)^{1/p}$, where the $p$th root is taken to be 
real and positive. The equation of motion for $\phi$ is
\be
{\phi}^{\prime\prime}(x) = {V_{p}}^{\prime} ({\phi}(x)),
\ee
where, the prime denotes derivative with respect to the argument. A classical
solution to this equation of motion is
\be
\bar\phi(x) = {\rm sech}^p(x/\sqrt{2p}),
\ee
which is a lump solution. 
If the field $\phi$ is expanded about the lump $\bar{\phi}$, the fluctuation
modes can be shown to satisfy the Schr\"{o}dinger equation\cite{Gold}
\be
-\frac{d^2 \psi(x)}{d x^2} + {U(x)}\psi(x) = m^2 \psi(x),
\ee
where ${U}(x) = V_{p}^{\prime \prime}(\bar{\phi}(x))$.
Using the form of the lump solution in this equation we can determine the form
of the quantum mechanical potential $U(x)$: 
\be
U(x) = \frac{1}{2p}\left(p^2 -
(p+1)(p+2)\rm{sech}^2(x/\sqrt{2p})\right).
\ee
This problem belongs to a class of exactly solvable quantum mechanical
models\cite{Khare}. For integer $p$, this potential admits $p+1$ bound states
and rest is a continuum with reflectionless scattering matrix. Using the
spectrum of this problem the mass spectrum of the fluctuations can be
deduced--- 
\be
m^2 = \frac{1}{2p}(p^2 - (p+1-n)^2), 0\leq n < p+1.
\ee
The lowest energy fluctuation mode is tachyonic with $m^2 = -1-{1\over 2p}$ 
and the next mode is massless.  In the limit $p\to\infty$, the tachyon
mass-squared becomes $m^2 =-1$, which mimics behaviour of the tachyon in 
the open
string theory. The potential in the field theory model when $p \to \infty$ 
reduces to
\be
V_{\infty}(\phi) = - A \phi^2 \ln(\phi^2).
\ee
This is a familiar potential in the background independent open string field
theory. 

The field theory problem with the potential $V_{\infty}$ admits a lump
solution which is a Gaussian and the Schr\"{o}dinger equation for the
fluctuation modes is that of the harmonic oscillator. The spectrum is
equispaced. The  lowest mode is tachyonic with $m^2 = -1$ and next mode is
massless. 

As mentioned in the introduction, our motivation for studying generalizations
of this problem is to enquire if the Gaussian nature of the lump solution and
consequently equispaced fluctuation spectrum, in $p\to\infty$ limit is a
universal feature or not. In the rest of the paper we will consider a more
general class of potentials $V_p(\phi)$, seek a solution to the classical
equation of motion and study the spectrum of the fluctuation modes. We will
concentrate specifically on the $p\to\infty$ limit of these potentials to see
if the fluctuation spectrum is equispaced.

\section{Generalized Potentials}

In this section we will consider more general potentials $V(\phi)$ and seek
soliton solutions to the equations of motion. In the first subsection, we will
address a simple issue related to the reflectionless potentials in the 
Schr\"{o}dinger equation which appeared in the model studied by
Zwiebach\cite{Zwie}. These quantum mechanical potentials are parametrized by
an integer $p$, 
\be
U(x) = p^2 - p(p+1)\rm{sech}^2x.
\ee
If $p$ is not integer then the corresponding quantum mechanical problem does
not have reflectionless scattering matrix but the problem is still exactly
solvable. In the following we demonstrate that the reflectionlessness of the
potential is not relevant in the $p\to\infty$ limit and we still end up with a
harmonic oscillator problem for the fluctuation modes of the soliton solution.

In the second subsection, we will consider more general potentials which are
all unbounded below. An interesting feature of all these potentials is that
each has a point of inflection at $\phi=1$. This leads to a novel type of
soliton solution which has no unstable mode despite potential being unbounded.

\subsection{Simple Extension of the Minahan-Zwiebach model}

We will start with a minimal generalization of the Minahan-Zwiebach model. 
Consider a potential of the form
\begin{equation}
V_{p,q}(\phi)  =  A \phi^2 p(1 - \phi^{q/p} ),
\end{equation}
where $p$ and $q$ are positive integers and $A$ is some arbitrary constant.
For any relatively prime integers $p$ and $q$, equation of motion for this
potential has a soliton solution
\be
\bar\phi(x) \sim {\rm sech}^{2p\over q}({qx\over \sqrt{2p}}).
\ee
The stability analysis of this soliton solution gives us a Schr\"{o}dinger
equation with the potential
\be
U(x) = {1\over{2p}}\left(p^2 - (p+q)(p+ {q \over 2})\,{\rm
sech}^2({qx\over{\sqrt{2p}}})\right).
\ee
This potential is not reflectionless. However, the Schr\"{o}dinger equation is
still exactly solvable\cite{Khare}.

As mentioned in the previous section we will be interested in $p \to \infty$
limit. In this limit, scalar field theory potential becomes
\begin{eqnarray}
V_{\infty ,q}(\phi) & = &  \lim_{p \rightarrow \infty} V_{p,q}(\phi) 
\nonumber \\
& = & A \phi^2 \lim_{n \rightarrow 0}\frac{(1 - \phi^{qn})}{n}\nonumber\\
& = & - A \phi^2 \ln(\phi^q)= - Aq \phi^2 \ln(\phi)\label{potinf}.
\end{eqnarray}

This potential is identical to that obtained by Minahan and Zwiebach except
that $A$ is replaced by $Aq/2$. This, however, does not affect any of their
results, {\em viz.} lump with a  Gaussian profile and equispaced fluctuation
spectrum, as $A$ can easily soak up the additional factor.  So the lesson we
learn from this is that $p\to\infty$ can really be reached in a continuous
manner. Reflectionless feature of the potential for the fluctuation modes is
of no consequence. 

\subsection{New unbounded potentials}

\begin{figure}[h]
\centerline{\epsfxsize 14.5cm
            \epsfbox{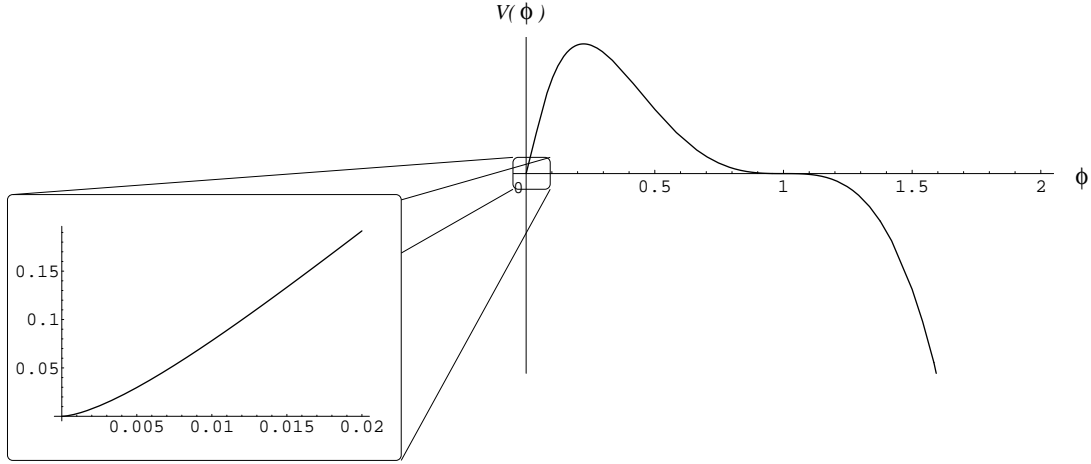}}
\caption{\small Generic form of $V(\phi)$ for odd $n$. The enlarged region
shows the behaviour near the origin.}\label{fig-pot}
\end{figure}

In this subsection we will consider more general potentials. Looking at the
original proposal of Zwiebach\cite{Zwie}, we see that at $p = \infty$ the
potential develops a simple zero, and is of indeterminate form $\infty \times
0$. The $p\to\infty$ limit is then determined using L'H\^{o}spital's rule. A
straightforward generalization of this is to consider potentials which have a
higher order zero as $p\to\infty$.  To this end let us consider
\be
V (\phi)  =  A \phi^2 p^n (1 - \phi^{q/p} )^n,
\ee
for an arbitrary odd integer $n > 1$. Recall that $n=1$ corresponds to the original
problem studied by Zwiebach, and Minahan and Zwiebach. As usual we will 
restrict ourselves to $p\to\infty$ limit. In this limit, the potential
becomes
\be
V (\phi) = A q^n \phi^2 (- \ln\phi)^n.\label{newpot}
\ee
Notice that for odd $n$, this potential is unbounded below.
The generic form of $V (\phi)$ for odd $n$ is shown in Fig.~\ref{fig-pot}. 
So far we have seen that a classical solution to the equation of motion with a
potential which is unbounded below has tachyonic mode. The potential
(\ref{newpot}) possesses a local minimum at $\phi=0$ where the first
derivative vanishes. It has a maximum at $\phi = e^{-n/2}$. At this unstable
point, $V^{\prime\prime}  = - 2A q^n (n/ 2)^{n-1}$, which is the mass-squared
of the tachyon.  There is also an inflection point at $\phi = 1$ at which the
first $n-1$ derivatives of the potential vanish. Since the potential, for odd
$n$, is unbounded below it is natural to expect that the lowest lying
fluctuation will destabilize the classical solution. However, we have a
surprise in store here. As we will see below, despite having an unbounded
potential, the classical solution that we get is not a lump solution but is a
kink. This is because of the fact that the potential (\ref{newpot}) has a
point of inflection at $\phi = 1$. 

A solution to the equations of motion can be obtained by solving the
integral equation
\be
\int \frac{d \bar{\phi}}{\bar{\phi} (- \ln \bar{\phi})^{n/2}}
= \pm \sqrt{2 q^n A} x,
\ee
and is given by
\beqn
& &\bar{\phi}_{\mp}(x) = \exp{\left[\mp \alpha_n
(x^2)^{-\frac{1}{n-2}}\right]},\quad \quad {\rm for}\, n\ge 3, \label{kink}\\
{\rm where}& & \quad \alpha_n = \left[\left(\frac{n}{2} - 1\right)^2 
              2 A q^n \right]^{-1/(n-2)}.
\eeqn
We will take $(n-2)$th root of $x^2$ to be real and positive. This is assured
if $n$ is an odd integer. Notice that there are two solutions corresponding to
each of the  $\mp$ signs in the exponent.  Let us first examine the solution
with the  positive sign, {\em i.e.}, $\bar{\phi}_+(x)$. This solution blows up
at $x=0$ and it approaches $\phi =1$ as $x\to\infty$. This field configuration
interpolates between the extremum at $\phi=\infty$ and the inflection point at
$\phi=1$. This solution is neither relevant nor accessible if we are studying
tachyon condensation, which corresponds to the field rolling down from the
maximum of the potential. The solution $\bar{\phi}_+(x)$ does not ever reach
the maximum of the potential. It always stays to the right of the inflection
point, while the maximum is to the left of the inflection point.

\begin{figure}
\epsfxsize 14.5cm
\epsfbox{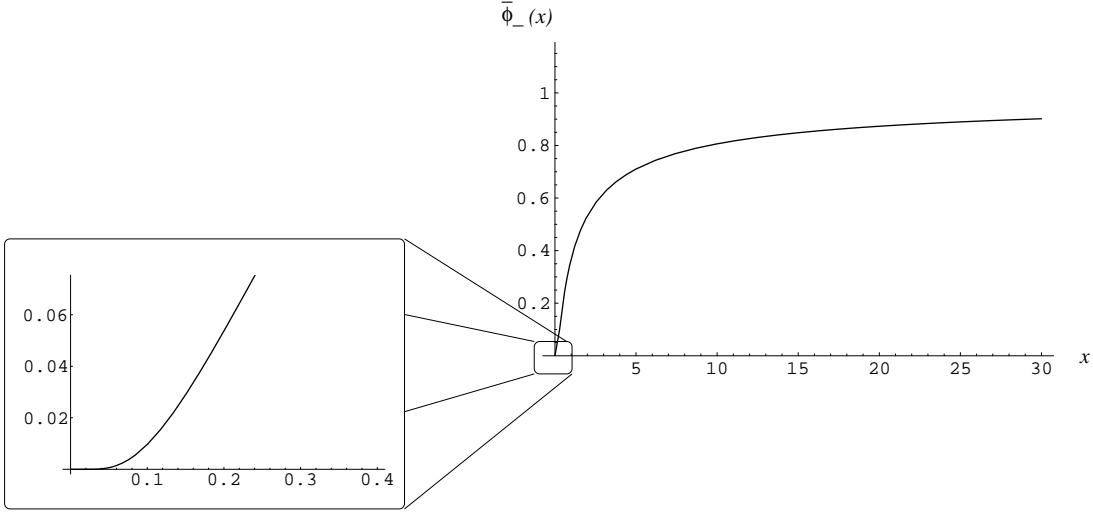}
\vskip 2mm
\caption{\small The kink $\bar{\phi}_-(x)$. The enlarged
region shows the behaviour near the origin.}
\label{fig-kink}
\end{figure}

The solution of interest to us is $\bar{\phi}_-(x)$, the profile of
which is shown in Fig.~\ref{fig-kink}. It approaches $\phi =0$ as
$x\to 0$, which is a local minimum of the potential and as $x\to\infty$ it
approaches $\phi =1$. For intermediate value of $x$, it rolls over the maximum
of the potential. At this point we would like to make a few remarks about the
solution $\bar\phi_-(x)$. As mentioned earlier, $\phi=1$ is a point of 
inflection and for all values of $n>2$, at least two derivatives of the 
potential with respect to $\phi$ vanish at that point. Using the equation of 
motion, this translates into the fact that at $\phi =1$ at least three 
derivatives of the solution $\bar\phi_-(x)$ with respect to $x$ vanish. Recall 
that the Lagrangian of our model contains terms which have at most two 
derivatives in it. Thus, there are no higher derivative terms in the theory 
which can destabilise this solution. We believe the situation is quite 
different if we view this problem from the string field theory perspective. The 
latter contains $*$-products of fields, which involves an arbitrary number of 
derivatives of the string field. These terms may destabilise the solution 
$\bar\phi_-(x)$. At this point, one may wonder why we are not looking at the
solution for negative values of $x$. The reason for this is that the kink
solution is such that as we approach $x=0$, $\bar\phi\to 0$ in such a way that
all its derivatives vanish at $x=0$, {\em i.e.}, the soliton solution has an
essential singularity at $x=0$. In the mechanical model analogy\cite{Raja},
the particle will take infinite time to reach $x=0$ from $x>0$ and hence $x<0$
region will not be reachable. We will come back to this point after we study
the mass spectrum of the fluctuations around this kink solution.

The potential which governs the mass spectrum of the fluctuation modes is
\beqn
U_n(x) = \frac{2}{(n-2)^2} \left[ 
                2 \alpha_n^2 \left(\frac{1}{x^2}\right)^{n/(n-2)}
              -3n \alpha_n \left(\frac{1}{x^2}\right)^{(n-1)/(n-2)}
          +n(n-1) \frac{1}{x^2} \right].
\eeqn
It is possible to recast the Schr\"{o}dinger equation with this potential in 
terms of supersymmetric quantum mechanics\cite{Khare}. The corresponding
quantum mechanical superpotential is given by
\beqn
W(x) & =&  \frac{n}{n-2} \frac{1}{x}
      - \left[ \frac{2^{n-1}}{(n-2)^n Aq^n}\right] ^{1/(n-2)}
           \frac{1}{x^{n/(n-2)}}\\
    &  =&  \frac{1}{n-2}\left[ \frac{n}{x} - 
\frac{ 2 \alpha_n}{x^{n/(n-2)}}\right].
\eeqn
This allows us to calculate the exact ground state wavefunction,
\beqn
\psi_0(x) & = & \exp{ \left( -\int W(x) dx \right)} \label{gs}\\
          & = & x^{-n/(n-2)} \, \exp{ \left( -\alpha_n x^{-2/(n-2)}
\right)}.
\eeqn
Exact supersymmetry of the Schr\"{o}dinger equation for the fluctuation
spectrum implies that the ground state energy is zero, {\em i.e.}, the lowest
lying fluctuation mode is massless. Thus, the fluctuation spectrum does not
contain a tachyon, which means that the kink solution is stable.

Another way of seeing that the kink solution is stable is the following.
The existence of a space-dependent solution spontaneously breaks translation 
invariance. Studying low energy excitations around that solution then gives the
massless Goldstone mode, which is responsible for restoring broken translation
invariance. If the solution is unstable, the lowest energy mode is tachyonic,
and the massless Goldstone mode is a higher level excitation. The Goldstone
mode for our kink solution can be deduced by looking at the first derivative
of the kink solution, $\bar\phi$ with respect to $x$.  It is straightforward
to see from Eq.(\ref{kink}) that $\bar\phi^\prime$ is proportional to the
ground state (\ref{gs}) of the fluctuations. As we argued above, this lowest
energy state, thanks to supersymmetry, has zero energy and hence the
corresponding fluctuation mode is massless.

\begin{figure}
\begin{center}
\epsfxsize 14.5cm
\epsfbox{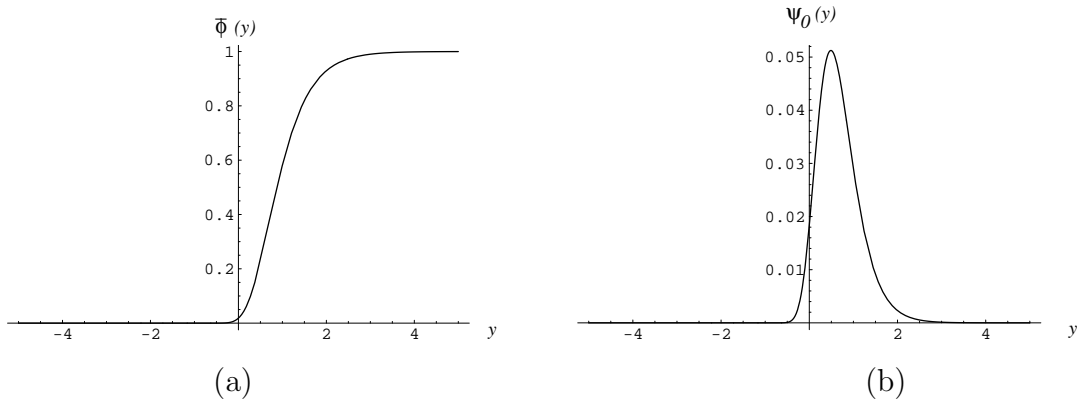}

(a) \hskip 8cm (b)
\end{center}
\caption{\small  Profiles of (a) the kink solution $\bar\phi$ and (b) the
ground state wavefunction $\psi_0$ of fluctuations, in the coordinate $y=\ln
x$ } \label{fig-exp}
\end{figure}

Let us come back to the point of soliton profile in the negative $x$
direction. The ground state wavefunction (\ref{gs}) is the lowest lying
fluctuation of the soliton, which as we just argued is massless. The profile
of the ground state wavefunction shares same features as that of the soliton
near $x=0$. That is, the ground state wavefunction vanishes as $x\to 0$ with
all its derivatives vanishing. The wavefunction also vanishes as $x\to\infty$.
If we extend the soliton solution beyond $x=0$ along the negative $x$ axis,
the lowest lying mode will develop a node at $x=0$, which is a contradiction.
The resolution is that $x$ is not a correct choice of coordinate. If we work
in terms of the coordinate $y=\ln x$ then we resolve both these issues.
The soliton profile and the ground state wave function in this
coordinate are  shown in Fig.~\ref{fig-exp}.

It is instructive to do the following field redefinition\cite{GS},
\be
\phi = \exp(-\chi).
\ee
With this field redefinition,
Lagrangian density of Minahan and Zwiebach model becomes
\be
{\cal L} =-{1\over 2} \exp(-2\chi)\left(\partial_{\mu}\chi\partial^{\mu}\chi  
+ \chi\right).\label{LBSFT}
\ee
Let us consider the same field redefinition for our problem. It is
straightforward to see the the Lagrangian density is
\be
{\cal L}_n =-{1\over 2} \exp(-2\chi)\left(\partial_{\mu}\chi\partial^{\mu}
\chi + 2Aq^n\chi^n\right).\label{MBSFT}
\ee
It is well known that (\ref{LBSFT}) is identical to the two derivative
reduction of the boundary string field theory action. It is tempting to
speculate that the Lagrangian density (\ref{MBSFT}) for our field theory
models may be related to the multicritical phenomena in the boundary string
field theory. It is also interesting to ask what the interpretation is of
these solitons in string field theory.

\section{Conclusions}

In this work, we have studied a single scalar field theory model with
potentials which are unbounded. We have obtained soliton solutions to the
equations of motion. A single most interesting and intriguing aspect of these
solutions is that they are stable, {\em i.e.}, their fluctuation spectrum does
not contain any tachyonic mode.

We have addressed the questions raised in the introduction, {\em viz,} whether
the fluctuation spectrum of lump solutions of generic potentials with a
parameter $p$, in the $p\to\infty$ limit, is equispaced with a tachyonic mode.
The answers turn out to be in the negative.  A simple extension of the
potential studied by Minahan and Zwiebach exhibits similar behaviour, {\em
viz.} the lump has a Gaussian profile and a tachyonic instability. The modes
are still equispaced and the mass-squared of the tachyonic mode is the same as
that of the fluctuation of the field about the potential maximum.  However,
the potential governing the fluctuation spectrum of the finite $p$ case is
{\it not} reflectionless, albeit of a similar class of exactly solvable
models.  The $p \to \infty$ limit can thus be taken in a continuous manner:
one does not need to restrict $p$ to be an integer. Thus, the property of
reflectionlessness of the potential is not crucial to the argument of Minahan
and Zwiebach.

For various other types of potentials, which are all unbounded, we find that
the Gaussian lump solution is not universal. In fact, we see that the soliton
solution is not even lump-like, it is instead a kink solution. We find that
the small fluctuation mass spectrum for these new solitons is not equispaced.
This is because the differential equation for the fluctuation modes is a
Schr\"{o}dinger equation with a potential which is not that of  harmonic
oscillator type. The tachyonic instability observed by Minahan and Zwiebach
for the Gaussian lump solution is also not universal. The kink solutions we
obtain are stable---there is no tachyon in the fluctuation spectrum. All the
potentials we have studied have one common feature, namely, a point of
inflection at $\phi=1$, where at least two derivatives of the potential
$V(\phi)$ with respect to $\phi$ vanish. The kink solution interpolates
between the local minimum at $\phi=0$ and the point of inflection $\phi=1$. It
approaches both these points infinitely slowly  and therefore field
configurations beyond $\phi=1$ are not accessible. This, we believe, is the
reason for non-existence of tachyon on the kink solution.

The new field theory models, with the fluctuation spectrum governed by a
supersymmetric quantum mechanical potential, are of interest in their own
right. The odd $n$ cases considered here have unbounded potentials, but the
even $n$ cases have bounded potentials. The latter belong to supersymmetric
field theories\cite{MZ1,H1}.  The soliton solution for these potentials has the
same profile (Eq.~\ref{kink}). The bosonic as well as fermionic zero-modes can
be calculated exactly\cite{JSV}. The relation of these potentials to tachyon
condensation in superstring field theory is under investigation.  Models with
finite  $p$ are also being analyzed. It would be interesting to see if they
have some surprise in store.

\vspace*{3mm}

\acknowledgments
{
We thank Ashoke Sen for asking a question which led to this investigation. 
We also thank him for discussion and critical reading of the manuscript. We 
thank Avinash Khare, Sumathi Rao and Subrata Sur for discussion. R.V. 
acknowledges support by CSIR, India, under grant no. 9/679(7)/2000-EMR-I.
}

\end{document}